\documentclass{INTERSPEECH2023}
\usepackage{multirow}
\usepackage[ruled]{algorithm2e}
\interspeechcameraready 

\title{An Intra-BRNN and GB-RVQ Based END-TO-END Neural Audio Codec}
\name{Linping Xu$^{2 \ast}$, Jiawei Jiang$^{1}$, Dejun Zhang$^{1}$, Xianjun Xia$^{1}$, Li Chen$^{1}$, \\Yijian Xiao$^{1}$, Piao Ding$^{1}$, Shenyi Song$^{1}$, Sixing Yin$^{2}$, Ferdous Sohel$^{3}$}
\address{
$^{1}$  RTC Lab, ByteDance, Beijing, China \\ 
$^{2}$ Beijing University of Posts and Telecommunications, Beijing, China \\
$^{3}$  School of Information Technology, Murdoch University, Perth, Australia
}
\email{
$^{1}$\{jiangjiawei.lahm, zhangdejun, xiaxianjun, chenli.cloud, dingpiao1, songshenyi1\}@bytedance.com 
$^{2}$\{yunzhongxue521, yinsixing\}@bupt.edu.cn, $^{3}$F.Sohel@murdoch.edu.au
}

\begin{document}

\maketitle

\renewcommand{\thefootnote}{\fnsymbol{footnote}}
\footnotetext[1]{This work was done when Linping Xu was an intern at ByteDance.} 
 
\begin{abstract}
Recently, neural networks have proven to be effective in performing speech coding task at low bitrates.  However, underutilization of intra-frame correlations and the error of quantizer specifically degrade the reconstructed audio quality. To improve the coding quality, we present an end-to-end neural speech codec, namely CBRC (Convolutional and Bidirectional Recurrent neural Codec). An interleaved structure using 1D-CNN and Intra-BRNN is designed to exploit the intra-frame correlations more efficiently. Furthermore,  Group-wise and Beam-search Residual Vector Quantizer (GB-RVQ) is used to reduce the quantization noise. CBRC encodes audio every 20ms with no additional latency, which is suitable for real-time communication. Experimental results demonstrate the superiority of the proposed codec when comparing CBRC at 3kbps with Opus at 12kbps.

\end{abstract}

\section{Introduction}
\label{Codec_intro}

Traditional audio codecs utilizing psycho-acoustics and pronunciation models are capable of producing high-quality audio at medium-to-high bitrates, whereas fail to produce satisfactory perceptual quality due to the inefficient coding architecture at low bitrates. With the development of deep learning, data-driven neural audio codecs provide a new direction for high-quality speech coding at low bitrates.

Neural networks used in audio codec can mainly be divided into three categories: post-processors, neural decoders and end-to-end neural codecs. Related works in \cite{zhao2018convolutional, korse2020enhancement, korse2022postgan} show that the quality of existing codecs can be improved by using neural network based post-processors without  changing the traditional codec's bitstream. Typical neural decoders are dependent on carefully designed audio features and generative models to reconstruct audio waveforms, such as WaveNet \cite{wavenet}, Lyra \cite{lyra}, LPCNet \cite{lpcnet}, SampleRNN \cite{samplernn} and SSMGAN \cite{streamwisestylemelgan}. They perform better than traditional codecs at low bitrates, especially with the use of GAN.

Different from post-processors and neural decoders, the learnable encoder in end-to-end neural codec significantly improves coding efficiency with respect to traditional codec. A learnable residual vector quantizer (RVQ) is adopted in SoundStream \cite{soundstream} which makes it the fully end-to-end codec.
SoundStream permits to encode audio at bitrates ranging from 3 kbps to 12 kbps with structured dropout applied to RVQ during training. Experimental results in that work demonstrate that Soundstream is robust under a wide range of real-life coding scenarios. Another end-to-end codec, TFNet \cite{tfnet}, takes temporal filtering blocks to explicit audio feature and investigates the joint optimization considering both speech enhancement and packet loss concealment task. More recently, NESC \cite{nesc} adopts the DPCRNN as the main building block for efficient and reliable encoding and demonstrates its robustness under various noise and reverberation levels.

Although SoundStream and NESC both achieve better quality than traditional codecs at low bitrates, the encoder and decoder in SoundStream adopt fully convolutional network, which pays little attention to intra-frame correlations. However the state-of-the-art traditional codecs, e.g. Opus and EVS, demonstrate the importance of intra-frame correlations.
Meanwhile, RVQ used in SoundStream and NESC provides a basis for end-to-end optimization. But there is still a large gap between the potential bitrate achievable by entropy coding and the actual bitrate of RVQ. 

To utilize the intra-frame correlations and reduce the quantionzation error, we propose CBRC, an end-to-end neural codec, capable of coding wideband audio at 3kbps. In encoder and decoder of CBRC, each Convolutional and Bidirectional Recurrent Neural Network Block (CBRNBlock)  adopts an interleaved network using 1D-CNN and intra-frame bidirectional RNN (Intra-BRNN)  to effectively capture intra-frame correlations. Meanwhile, Group-wise and Beam-search RVQ are designed in quantizer to reduce the quantization error. This paper compares the quality of audio generated by our proposed CBRC, Lyra-V2 and Opus between 3kbps and 12kbps. It is worth noting that CBRC at 3kbps outperforms Opus at 12kbps.

Our contributions are listed as below: i) The model architecture with Intra-BRNN and ii) The Group-wise RVQ and Beam-search RVQ methods adopted during quantization, which are the main aspect of novelty of our proposed CBRC. 


\section{Proposed Audio Codec}
\label{CBRC_codec}
\begin{figure}[t]
  \begin{minipage}[b]{1.0\linewidth}
    \centering
    \centerline{\includegraphics[width=0.9\linewidth,height=0.45\linewidth]{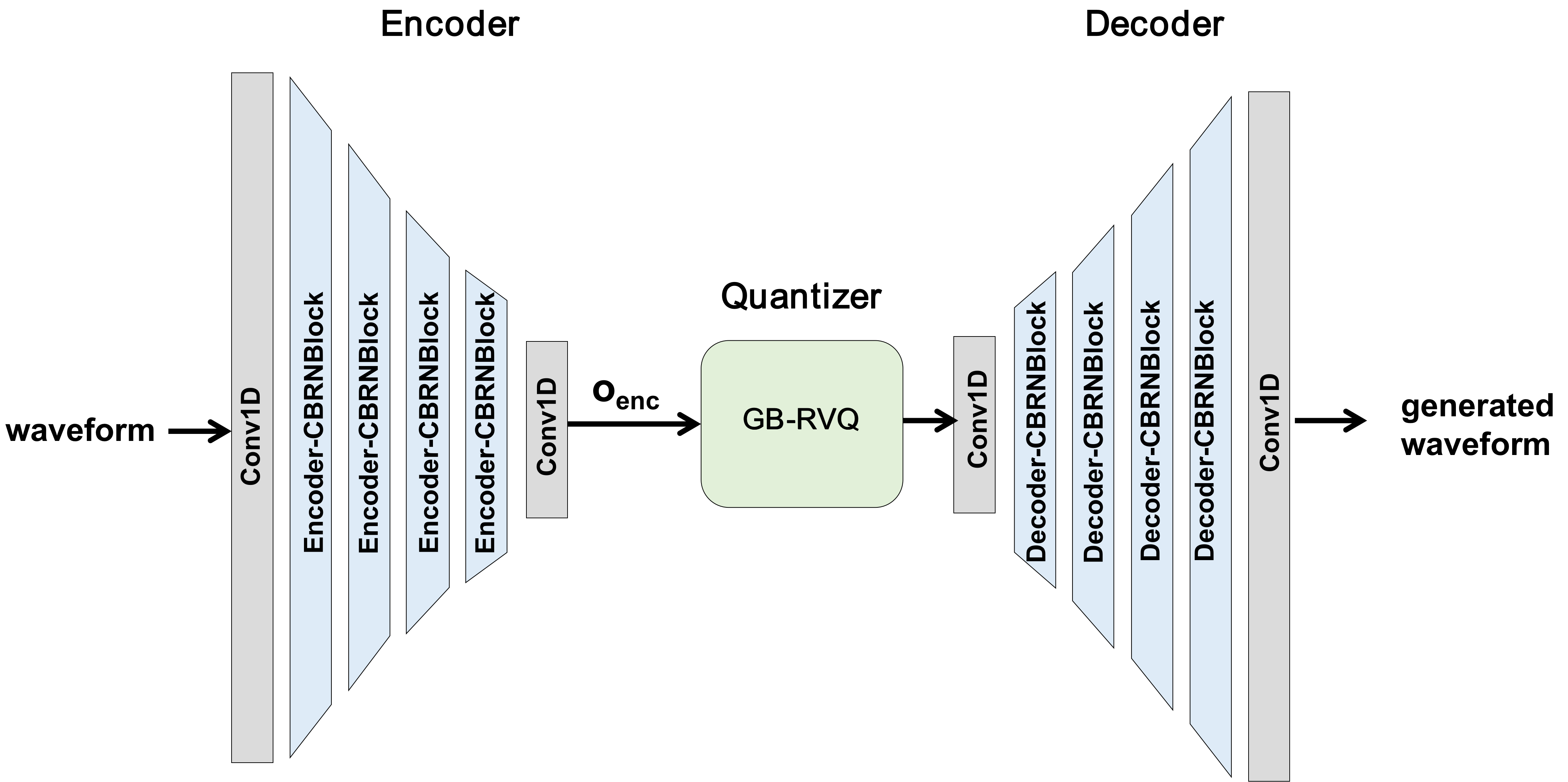}}
  \end{minipage}
  \caption{
    CBRC model architecture 
  }
  \label{cbrc_codec}
  \vspace{-0.2cm}
  \end{figure}

Figure \ref{cbrc_codec} shows the flowchart of our proposed codec, which consists of an encoder, a quantizer and a decoder. The encoder maps audio inputs to a sequence of embeddings. The quantizer performs to compress the embeddings with a target number of bits. The decoder is adopted to reconstruct audio from quantized embeddings. The codec is trained end-to-end with discriminator and adversarial loss to improve the perceptual quality of the reconstructed audio.

\subsection{Encoder and Decoder}
The encoder adopts four cascaded CBRNBlocks to perform the feature extraction. Each CBRNBlock is composed of three ResidualUnits for extracting features and a Conv1D layer with stride for down-sampling. Same as SoundStream \cite{soundstream}, the number of channels is doubled after down-sampling. Supposing the audio waveform $ x \in \mathbb{R}^{T}$ is sampled at $f_s$ with a duration of $T$, the encoder outputs can be expressed as:
\begin{eqnarray}\label{eq1}
\mathbf{o}_{\text{enc}} = &\text{Encoder}(\mathbf{x}), \mathbf{o}_\text{enc} \in \mathbb{R}^{S\times D}\\
&S = T / M
\end{eqnarray}
where $D$ is the encoder embedding dimension, $M$ denotes the sub-frame hopsize and $S$ means the number of embeddings. The decoder mirrors the encoder, using a transposed convolution for up-sampling. 

\begin{figure}[t]
  \begin{minipage}[b]{1.0\linewidth}
    \centering
    \centerline{\includegraphics[width=1\linewidth]{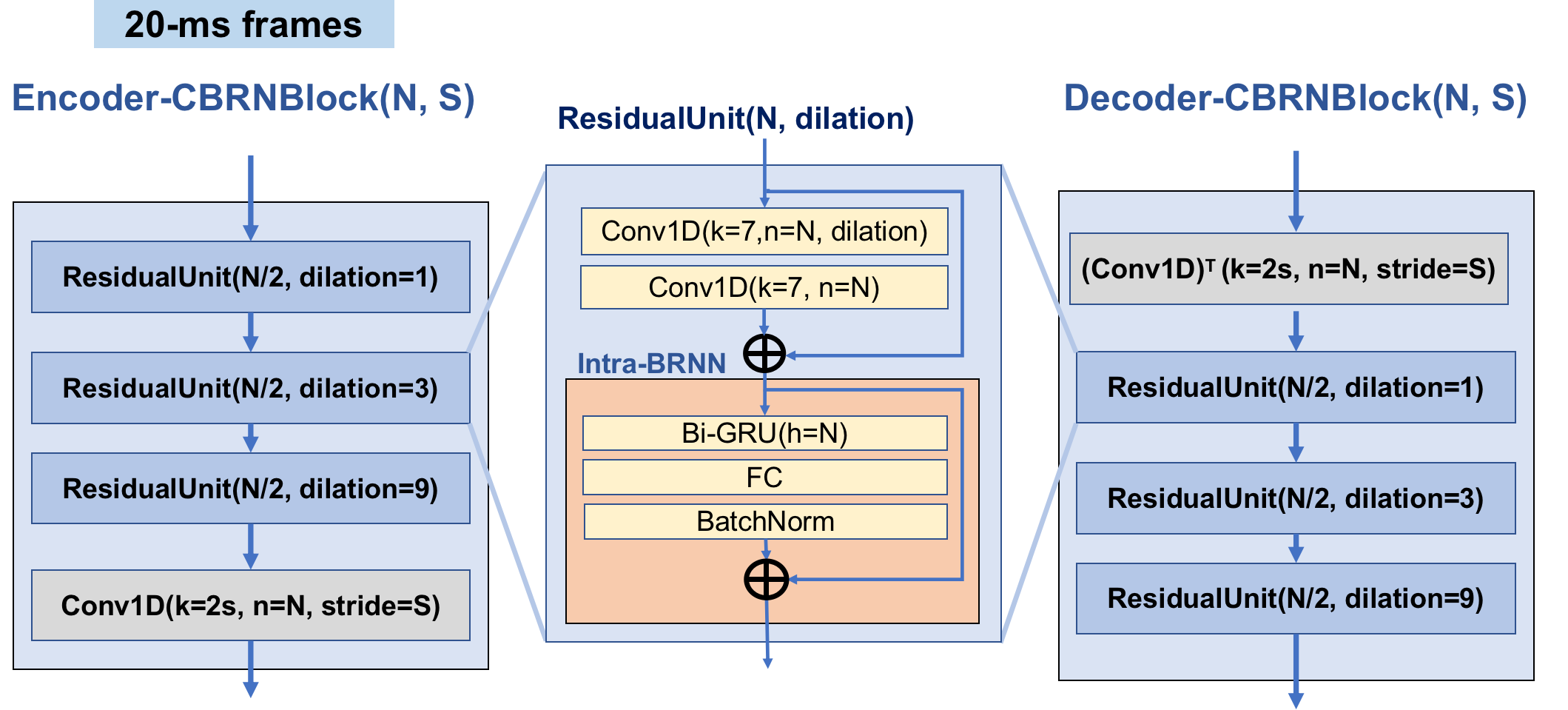}}
  \end{minipage}
  \caption{
    CBRNBlock diagram  
  }
  \label{encoder}
  \vspace{-0.2cm}
  \end{figure}

Figure \ref{encoder} illustrates the architecture of the CBRNBlock. Each ResidualUnit stacks dilated causal 1D-CNN and Intra-BRNN to form an interleaved structure. An internal buffer is inlcuded in each convolution to use information from past frames in inference mode. 

Intra-BRNN which consists of Bi-GRU, followed by a linear fully-connected layer and a batchnorm layer, is used in every 20ms frame to capture the intra-frame correlations with no additional delay.   

\subsection{Quantizer: Group-wise RVQ}
The goal of the quantizer is to compress the encoder output $\mathbf{o}_{\text{enc}} $ to a target bitrate $R$, expressed in bits/second (bps). A codeword is selected from $N$ vectors in Vector Quantizer (VQ) to encode embeddings. Embeddings $\mathbf{o}_{\text{enc}} $  are then mapped to a sequence of one-hot vectors of shape $(S, N)$ , which can be represented using $Slog_{2}N$ bits. As can be calculated, the codebook size of plain VQ is huge and computation complexity is also very high \cite{soundstream}.

Residual Vector Quantizer (RVQ) \cite{soundstream} cascades several VQ layers to reduce the codebook size. Each RVQ layer takes the quantization residual from the previous layer as the layer input. The codebook size of each quantizer can be calculated as with $N_q$ denoting the number of VQ layers: 
\begin{equation}
N = 2^{R/N_{q}/S}
\end{equation}

\begin{figure}[t]
  \begin{minipage}[b]{1.0\linewidth}
    \centering
    \centerline{\includegraphics[width=\linewidth,height=0.6\linewidth]{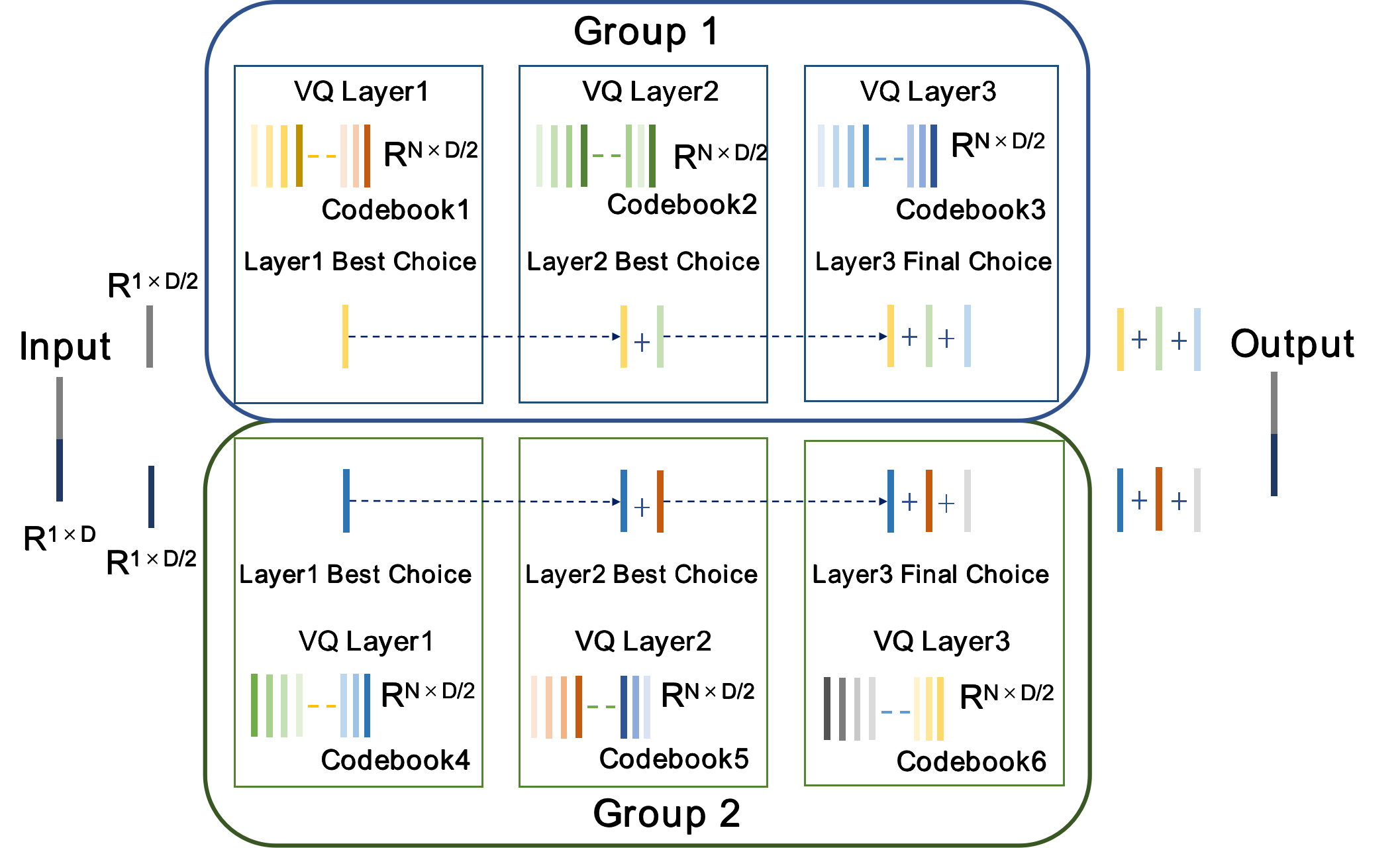}}
  \end{minipage}
  \caption{ Group-wise RVQ: The encoded embedding is split into $G$=2 independent embeddings. They are processed in parallel in quantizer. }
  \label{g_rvq}
\end{figure}

Figure \ref{g_rvq} shows  a more promising quantization architecture, namely the Group-wise RVQ. Group-wise RVQ splits the $D$-dimensional encoder output into $G$ sub-groups ($G$ is set to 2 in Fig. \ref{g_rvq} for demonstration). Each sub-group uses a smaller ($D/G$) codebook size and quantizes the split embedding independently. Afterwards, $G$ parallel optimal codewords from sub-groups are concatenated to form the final quantization result.  The number of parameters in RVQ codebook and the computational complexity of RVQ can be expressed as:
\begin{equation} \label{eq3}
\begin{split}
\text{Number of RVQ parameters}= N_q \times D \times N \\
\text{Complexity} =  N_q \times (D+1) \times N \times S
\end{split}
\end{equation}
As can be calculated, the number of RVQ parameter and quantization complexity reduced from 1.57M and 0.158 Macs to 0.79M and 0.079 Macs with Group-wise RVQ ($G$=2), in which $R$ is set to 6kpbs, $D$ is set to 256 and $S$ equals 100 when 10ms sub-frame hopsize is used within a 1-second audio segment.  Although the number of  parameters and complexity are halved, Group-wise RVQ outperforms the typical RVQ by VISQOL score 0.11(see Section 3.4).  

\subsection{Quantizer: Beam-search RVQ}
To reduce the quantization error, Beam-search algorithm based RVQ is proposed to select codebook vectors in a larger search space. Algorithm \ref{beam-search} shows Beam-search implementation details. As can be seen in Algorithm \ref{beam-search}, Beam-search RVQ selects vectors based on the minimum quantization path error measured by the mean square error (MSE) between the sum of selected vectors and the input embedding. The selection process can be divided into:
\textbf{1)} Each VQ layer quantizes the potential $k$ candidates leading to $k^{2}$ new options, \textbf{2)} Top $k$ paths with smaller errors are then selected among the  $k^{2}$ options as optimal outputs for current VQ layer and \textbf{3)}The best path is determined in the last VQ layer. 

\begin{figure}[t]
  \begin{minipage}[b]{1.0\linewidth}
    \centering
    \centerline{\includegraphics[width=1\linewidth,height=0.5\linewidth]{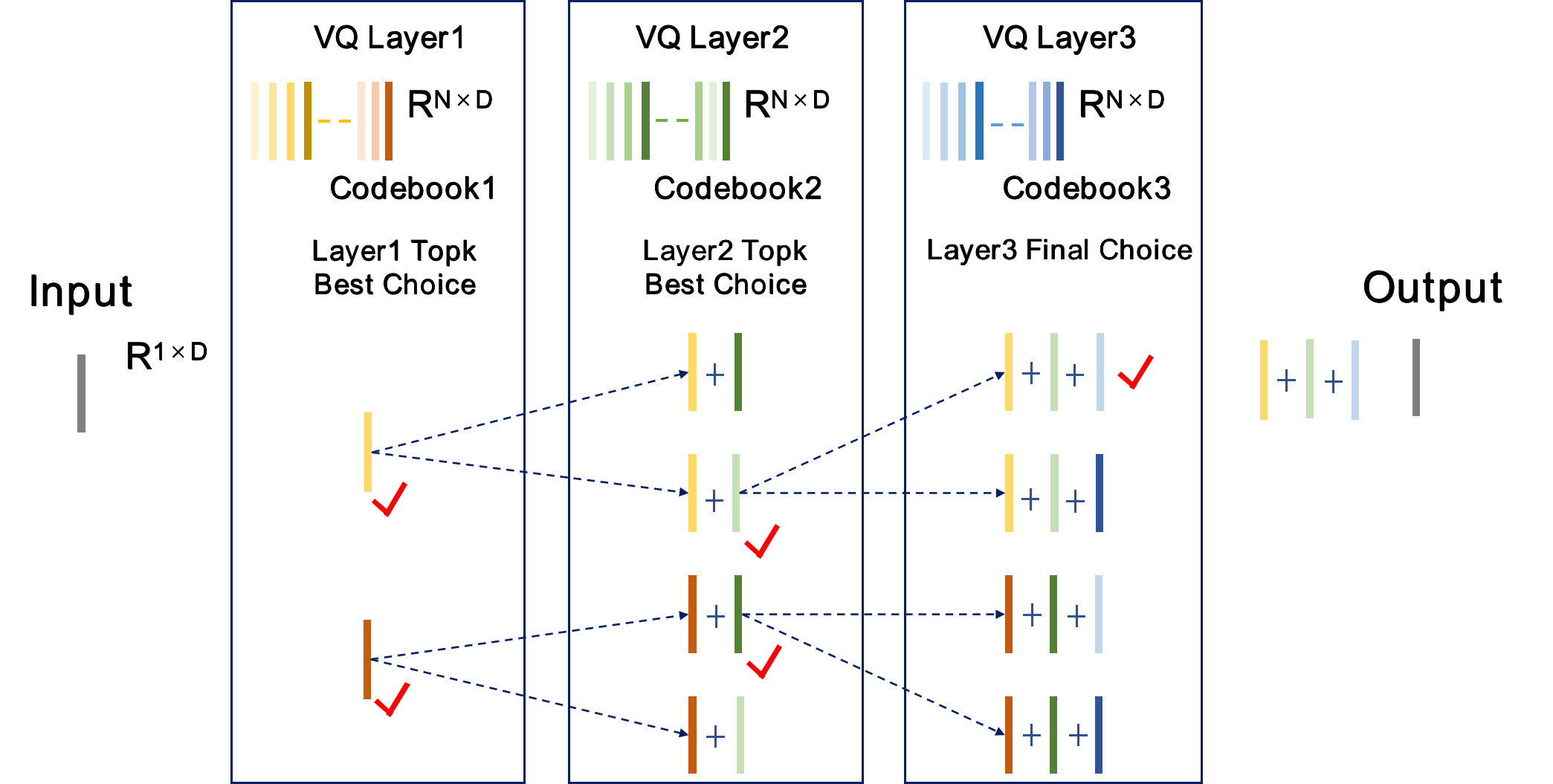}}
  \end{minipage}
  \caption{Beam-search RVQ: The figure shows the process of quantization in three layers of VQ when the number of candidates $k$ = 2. Each layer of VQ except the last reserves 2 quantization paths.  
  }
  \label{bs_rvq}
  \vspace{-0.3cm}
  \end{figure}

Figure \ref{bs_rvq} shows an example when $k$ is set to 2. A larger $k$ in Beam-search RVQ is associated with a larger search space and a smaller quantization error. Beam-search RVQ slightly increases by about $(1+k(N_{q}-1))/ N_{q}$ times with $N_q$ VQ layers and $k$ candidates.

\begin{algorithm}[t] 
  \caption{Beam-search RVQ}
  \LinesNumbered 
  \label{beam-search}
  \KwIn{
    Embedding X; Number of candidates k; 
    Vector Quantizers $Q_{i} $  for i = 1 ... $N_{q}$, where $\left[ z_{1}, ... , z_{k}\right] = Q_{i}(residual)$;
  $Res_{topk}= \left[ r_{1}, ... , r_{k}\right]$ where $r_{i}=X$;
  $\left[s_{1}, ..., s_{k}\right] = find(Candidates, k)$ where pick the k values with the smallest value }
  \KwOut{Quantized output Y}
  \For{i= 1 to $N_q$}{
    $Res\_tmp=\left[ \quad \right]$\;
    \For{j = 1 to k}{
      $r = Res_{topk}\left[j\right]$\;
      $\left[ z_{1}, ... ,z_{k}\right] = Q_{i}(r)$\;
      $\left[ r_{1}, ... ,r_{k}\right] = \left[ r-z_{1}, ... ,r-z_{k}\right]$\;
      $Res\_tmp.append(\left[ r_{1}, ..., r_{k}\right])$\;
      
    }
    $Res_{topk} = find(Res\_tmp, k)$
   
  }
  $Y = find(Res_{topk}, 1) $


\end{algorithm}

\subsection{Loss function}
The discriminator of the experiments is the same as the discriminator of SMGAN \cite{stylemelgan}. 
The overall function $\mathcal{L}$ can be expressed as:
\begin{equation}
  \begin{split}
    \mathcal{L}= &  \mathcal{L}_{\text{rec}}+ \lambda_{\text{feat}}\mathcal{L}_{\text{feat}}  + \lambda_{\text{adv}}\mathcal{L}_{\text{adv}} \\
    & +  \lambda_{\text{VQ}}\mathcal{L}_{\text{VQ}}+ \lambda_{\text{pmsqe}}\mathcal{L}_{\text{pmsqe}}
  \end{split}
  \end{equation}
where $\mathcal{L}_{\text{rec}}$ is the multi-scale reconstruction loss  \cite{soundstream}, $\mathcal{L}_{\text{feat}}$ is computed in the feature space defined by the discriminators \cite{soundstream}, $\mathcal{L}_{\text{adv}}$ is the adversarial loss for the generator, $\mathcal{L}_{\text{VQ}}$ is the commitment loss \cite{VQVAE}, which constraints vector quantization and $\mathcal{L}_{\text{pmsqe}}$ is the perceptual loss \cite{pmsqe_loss}, which is designed to be inversely proportional to PESQ \cite{pesq}. In this work, $\lambda_{\text{feat}}$, $\lambda_{\text{adv}}$, $\lambda_{\text{VQ}}$ and $\lambda_{\text{pmsqe}}$ are set to 100, 1, 1, 1, respectively.

\section{Experiments}
\subsection{Datasets and evaluation metrics}
CBRC was trained on 245 hours of speech from the LibriTTS dataset \cite{libritts} at 16 kHz. To ensure that the model is robust with different audio amplitudes, each segment was normalized with a maximum peak value of 0.95 and multiplied by a random gain from 0.3 to 1. The model is trained on A100-SXM-80GB with a batch size of 60. The computational complexity of the CBRC is approximately 4.57G Macs and the total number of model parameters is approximately 4.38M. 
In this paper, ViSQOL V3 \cite{visqol} and PESQ \cite{pesq} are adopted to evaluate the objective quality of audio reconstructed from CBRC. For subjective test, we used unseen audio samples with a MUSHRA-inspired crowd-sourced method \cite{mush}. 

\subsection{Comparison with other codecs}

\begin{figure}[t]
  \begin{minipage}[b]{1.0\linewidth}
    \centering
    \centerline{\includegraphics[width=1.0\linewidth]{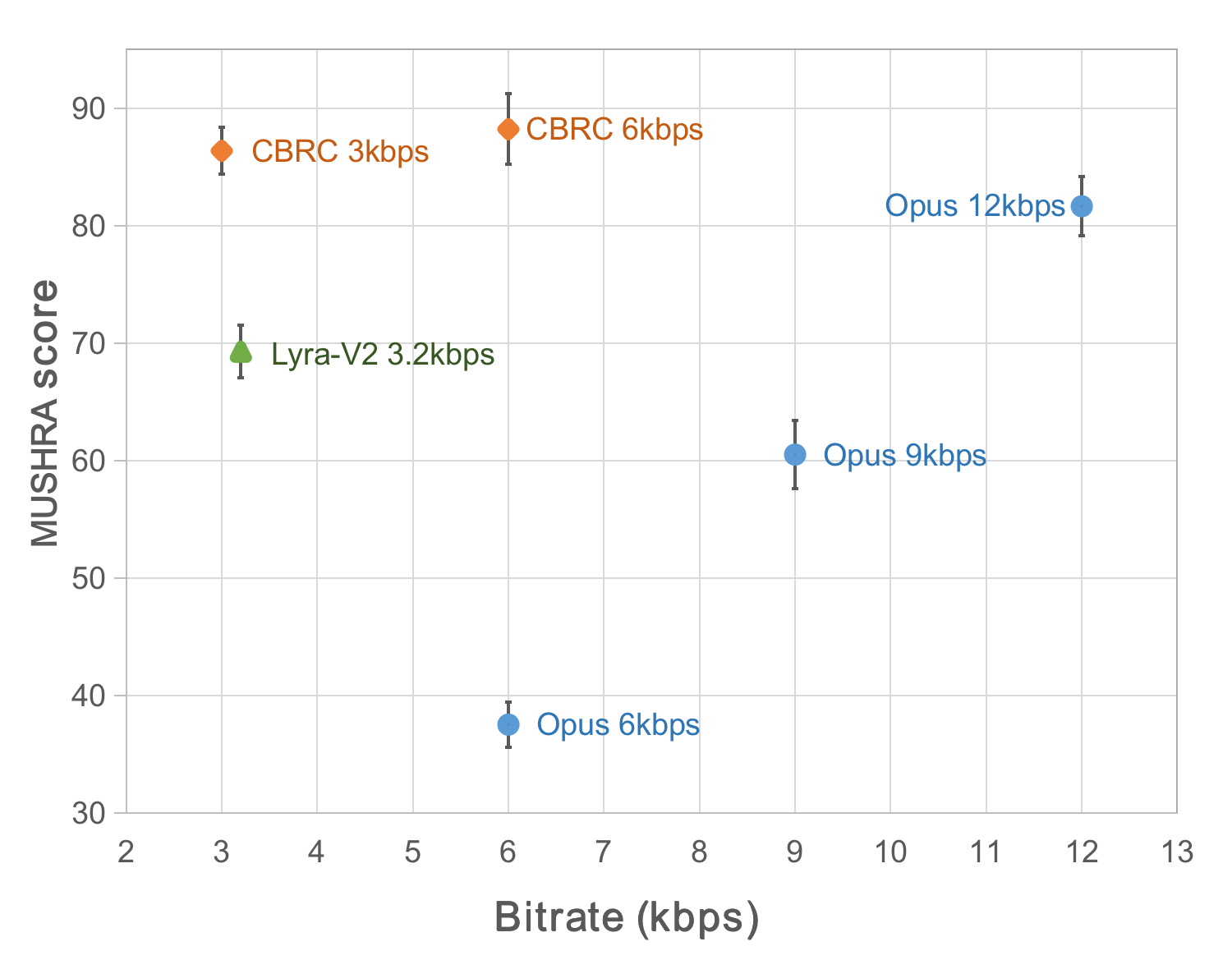}}
  \end{minipage}
  \setlength{\abovecaptionskip}{-0.4cm}
  \caption{Subjective scores for different codecs. Error bars denote 95\% confidence intervals.}
  \label{fig:subjective_score}
  \vspace{-0.3cm}
  
  \end{figure}

To evaluate the speech quality of CBRC, different codecs were evaluated on a set of 10 multilingual speech sequences. The subjective and objective scores are shown in Figure \ref{fig:subjective_score} and Table \ref{tab:different_codecs}. As can be seen from the results, our proposed codec at 3kbps outperforms Opus\footnote{https://opus-codec.org} at 12kbps, despite using a quarter of the bitrate.  

We also see that CBRC at 3kbps significantly outperforms Lyra-V2\footnote{https://github.com/google/lyra} at 3.2kbps, which demonstrates the superiority of the CBRC architecture. 
And in order to minimize the influence of complexity, we propose the lite version of CBRC with Intra-BRNN and Beam-search residual quantizer, which slightly increases complexity by 10\% but improves PESQ performance from 3.11 to 3.22 at 9kbps. The audio samples\footnote{https://bytedance.feishu.cn/docx/OqtjdQNhZoAbNoxMuntcErcInmb} are available.

\begin{table}[t]
  \caption{Objective scores of different codecs.}
  \label{tab:different_codecs}
  \vspace{-6mm}
  \begin{center}
    \begin{tabular}{lcccr}
      \textbf{Codec}     & \textbf{bitrate} & \textbf{Complexity} & \textbf{ViSQOL} & \textbf{PESQ} \\
                          & \textbf{}        & \textbf{(Macs)}     & \textbf{}       & \textbf{}     \\ \hline
      Opus               & 6kbps            &                     & 2.57            & 2.18          \\
      Opus               & 9kbps            &                     & 3.51            & 3.08          \\
      Opus               & 12kbps           &                     & 3.91            & 3.81          \\ \hline
      Lyra-V2            & 3.2kbps          & $\sim$344M          & 3.18            & 2.34          \\
      Lyra-V2            & 6kbps            & $\sim$344M          & 3.55            & 2.92          \\
      Lyra-V2            & 9.2kbps          & $\sim$344M          & 3.70            & 3.11          \\ \hline
      \textbf{CBRC}      & 3kbps            & 4.5G                & \textbf{3.71}   & \textbf{2.88} \\
      \textbf{CBRC}      & 6kbps            & 4.5G                & \textbf{4.04}   & \textbf{3.57} \\ \hline
      \textbf{CBRC-lite} & 3kbps            & 379.3M              & \textbf{3.56}   & \textbf{2.67} \\
      \textbf{CBRC-lite} & 6kbps            & 379.3M              & \textbf{3.84}   & \textbf{3.09} \\
      \textbf{CBRC-lite} & 9kbps            & 379.3M              & \textbf{4.01}   & \textbf{3.22} \\ \hline
      \end{tabular}

  \end{center}
  \end{table}

      


\subsection{Ablation of Intra-BRNN}

Several ablation experiments were carried out to evaluate the benefit of Intra-BRNN applied in CBRC. All the models were operated at 6kbps. Intra-BRNN can be applied in encoder, decoder or both. As can be seen in Table \ref{tab:dprnn_ablation}, the Intra-BRNN significantly improves ViSQOL from 3.41 to 4.04 when adopted in both encoder and decoder. To demonstrate the efficiency of codebook utilization in the proposed RVQ, the codeword frequency is counted in each VQ layer and the potential bitrate is calculated by entropy coding. The bitrates of the fully convolutional model and CBRC are 4.94kbps and 5.13kbps, which means the codebook utilization efficiency increased from 82.3\% to 85.5\% when target bitrate $R$ is to 6kbps. 

\begin{table}[t]
  \caption{Ablation of Intra-BRNN in encoder and decoder. $w/o$ indicates that the fully convolutional network, and $w$ indicates that the interleaved network of 1D-CNN and Intra-BRNN.}
  \label{tab:dprnn_ablation}
  \vspace{-4mm}
  \begin{center}
    \begin{tabular}{c|c|c|c}
      \hline
        & Encoder   & Decoder   & \multirow{2}{*}{ViSQOL} \\
        & Intra-BRNN & Intra-BRNN &                         \\ \hline

        1 & $w/o$        & $w/o$        & 3.41                     \\
        2 & $w/o$       & $w$        & 3.55                    \\
        3 & $w$        & $w/o$        & 3.88                    \\
        4 & $w$       & $w$       &  \textbf{4.04}                      \\\hline

        \end{tabular}

  \end{center}
  \vspace{-0.4cm}
  \end{table}

\begin{table}[t]
  \caption{Comparison of different types of RVQ.}
  \label{tab:rvq}
  \vspace{-4mm}
  \begin{center}
    \begin{tabular}{c|c}
      Quantizer Type & ViSQOL  \\ \hline
      RVQ            &     3.81     \\
      Group-wise RVQ      &   3.92     \\
      Beam-search  RVQ      &   4.02      \\
      GB-RVQ    &  \textbf{4.04}    \\ \hline          
      \end{tabular}
  \end{center}
  \vspace{-0.5cm}
  \end{table}

\subsection{Ablation of Group-wise RVQ and Beam-search RVQ}

To demonstrate the effectiveness of the proposed Group-wise RVQ and Beam-search RVQ,  different RVQs were evaluated on ViSQOL at 6kbps. Table \ref{tab:rvq} shows that the ViSQOL score improves  from 3.81 to 3.92 and 4.02 when the Group-wise and Beam-search algorithm are used, respectively. The Group-wise RVQ (number of sub-groups $G$=2) improves the performance and reduces the computational complexity. Beam-search RVQ (number of candidates $k$=4) slightly increases complexity of the quantization but significantly improves coding quality.  With Group-wise and Beam-search algorithm, the performance significantly increased from 3.81 to 4.04 and the computational complexity of quantizer slightly increases from 0.158G Macs to 0.277G Macs, which matters little compared to the whole model structure, complexity of which is 4.57G Macs.

\section{Conclusion}
\label{CBRC_conclusion}

In this paper, we propose an end-to-end neural audio codec, namely CBRC. An interleaved network of 1D-CNN and intra-BRNN were designed to fully capture temporal information. Furthermore, RVQ with Group-wise and Beam-search algorithms were developed to make the model perform better in neural codecs. The subjective and objective results show that CBRC achieves the state-of-the-art coding quality.  How to further reduce the coding redundancy and the computational complexity for real-time communication will be our future direction.

\bibliographystyle{IEEEtran}
\bibliography{CBRC}

\end{document}